\documentclass{aa}  
\usepackage{graphicx}
\usepackage{amsmath}
\usepackage{amsfonts}
\usepackage{amssymb}
\usepackage{gensymb}
\usepackage{textcomp}
\usepackage[varg]{txfonts}
\usepackage{natbib}
\usepackage[colorlinks,linkcolor=blue,linktocpage=true,plainpages=false,
urlcolor=blue,citecolor=blue,hypertexnames=false,breaklinks]{hyperref}

\begin{document}

   \title{A closer look at a coronal loop rooted in a sunspot umbra}

   \author{L.~P.~Chitta\inst{1}, H.~Peter\inst{1}, \and P.~R.~Young\inst{2,}\inst{3}}

   \institute{Max-Planck Institute for Solar System Research (MPS), 37077, G\"{o}ttingen, Germany\\
              \email{chitta@mps.mpg.de} 
              \and 
              College of Science, George Mason University, 4400 
              University Drive, Fairfax, VA 22030, USA
              \and
              NASA Goddard Space Flight Center, Solar Physics Laboratory, Greenbelt, MD
              20771, USA}

   \date{Received ; accepted }

 
\abstract
{Extreme UV (EUV) and X-ray loops in the solar corona connect regions of enhanced magnetic activity, but  they are {not} usually rooted in the dark umbrae of sunspots because the strong magnetic field found there suppresses convection. This means that  the Poynting flux of magnetic energy into the upper atmosphere is not significant within the umbra as long as there are no light bridges or umbral dots.}
%
{Here we report a rare observation of a coronal loop rooted in the dark umbra of a sunspot without  
any traces of light bridges or umbral dots. This allows us to investigate the loop without much 
confusion from background or line-of-sight integration effects.}
%
{We used the slit-jaw images and spectroscopic data from the Interface Region Imaging Spectrograph  
(IRIS) and concentrate on the line profiles of \ion{O}{iv} and \ion{Si}{iv} that show persistent 
strong redshifted components in the loop rooted in the umbra. Using the ratios of \ion{O}{iv}, we 
can estimate the density and thus investigate the mass flux. The coronal context and 
temperature diagnostics of these observations is provided through the EUV channels of the 
Atmospheric Imaging Assembly (AIA).}
%
{The coronal loop, embedded within cooler downflows, hosts supersonic downflows. The speed of more than 100\,km\,s$^{-1}$ is on the same order of magnitude in the transition region lines of 
\ion{O}{iv} and \ion{Si}{iv}, and is even seen at comparable speed in the chromospheric Mg {\sc ii} 
lines. At a projected distance of within $1''$ of the footpoint, we see a shock transition to smaller downflow speeds of about 15\,km\,s$^{-1}$ being consistent with mass conservation across a stationary isothermal shock.}
%
{We see no direct evidence for energy input into the loop because the loop is rooted in the dark  
uniform part of the umbra with no light bridges or umbral dots near by. Thus one might conclude that 
we are seeing a siphon flow driven from the footpoint at the other end of the loop. However, for a final 
result  data of similar quality at the other footpoint are needed, but this  is too far away to be 
covered by the IRIS field of view.
}
  
   \titlerunning{Coronal loop in a sunspot umbra}
   \authorrunning{Chitta et al.}
   \maketitle

%

\section{Introduction\label{intr}}
High-resolution imaging and spectroscopy of extreme UV (EUV) emission in solar coronal loops offer 
key diagnostics to understand how magnetic energy is converted into thermal energy. Diagnostics 
such as the Doppler velocity, density, and temperature of the plasma are of particular 
interest as they reveal the thermal evolution of coronal loops~\citep[for a recent review on 
various aspects of coronal loops and 3D modeling, see][]{lrsp-2014-4,2015RSPTA.37350055P}.
Over the last decade imaging and spectroscopic observations in combination with magnetograms 
have provided information on the coupling of coronal loops to the underlying magnetic 
field~\citep[e.g.,][]{2003A&A...406.1089D,2004A&A...428..629M} and have improved our understanding of 
coronal loops and flows within these loop structures. In the transition region (TR) from the 
chromosphere to the corona it has been observed that the footpoints of 1\,MK 
active region loops have strong emission at ${\log}T\,$[K]\,$=$\,5.4 to 5.8 and an electron density ($n_e$) 
of about $10^9~\text{cm}^{-3}$~\citep[][]{2003A&A...406L...5D,2007PASJ...59S.727Y}, while $n_e$ of active regions in the TR typically exceeds $10^{10}~\text{cm}^{-3}$ \citep[e.g.,][]{1997ApJ...476..903D}. Other well studied features of active region loops are pervasive and persistent downflows in the 
TR emission lines, and upflows in the hotter EUV 
lines~\citep[][]{2008A&A...481L..49D,2009ApJ...694.1256T}. \citet{2008ApJ...678L..67H} reported 
enhanced non-thermal velocities near the footpoints of active region loops, and recently 
\citet{2015ApJ...800..140G} derived properties of a coronal loop along its length. However,  the 
studies cited here -- and in fact most coronal studies -- suffer from line-of-sight confusion, 
background emission, and moderate spatial resolution, which completely limit loop and footpoint 
diagnostics in the TR and corona. It is highly desirable to have observations of EUV emission from 
isolated coronal loops and associated footpoints to place constraints on the loop heating models. 
One example is  the investigation of the cooling of a rather isolated loop in the periphery of an 
active region by \cite{2009ApJ...695..642U}. In contrast, finding an isolated footpoint is much more difficult.

In the light of the above discussion, we searched the database of the Interface Region Imaging 
Spectrograph \citep[IRIS;][]{2014SoPh..289.2733D}, which provides unprecedented observations of 
the chromosphere and TR at ${\approx}0''.33$ resolution. We looked for isolated footpoints and their loops -- which are in fact difficult to pinpoint -- and we identified isolated coronal loops and their footpoints that are rooted in the center of the observed umbra, which is an unusual place to find them. The theoretical reasoning why coronal loops should not be rooted in the center of the umbra is based on the simple argument that the strong field in the umbra suppresses any significant horizontal motions and thus the vertical Poynting flux into the upper atmosphere is very small there, which is confirmed in 3D magnetohydrodynamic models \cite[e.g.,][]{2014A&A...564A..12C,2015NatPh..11..492C}.

\begin{figure*}[!htb]
\centering
\includegraphics[width=17cm]{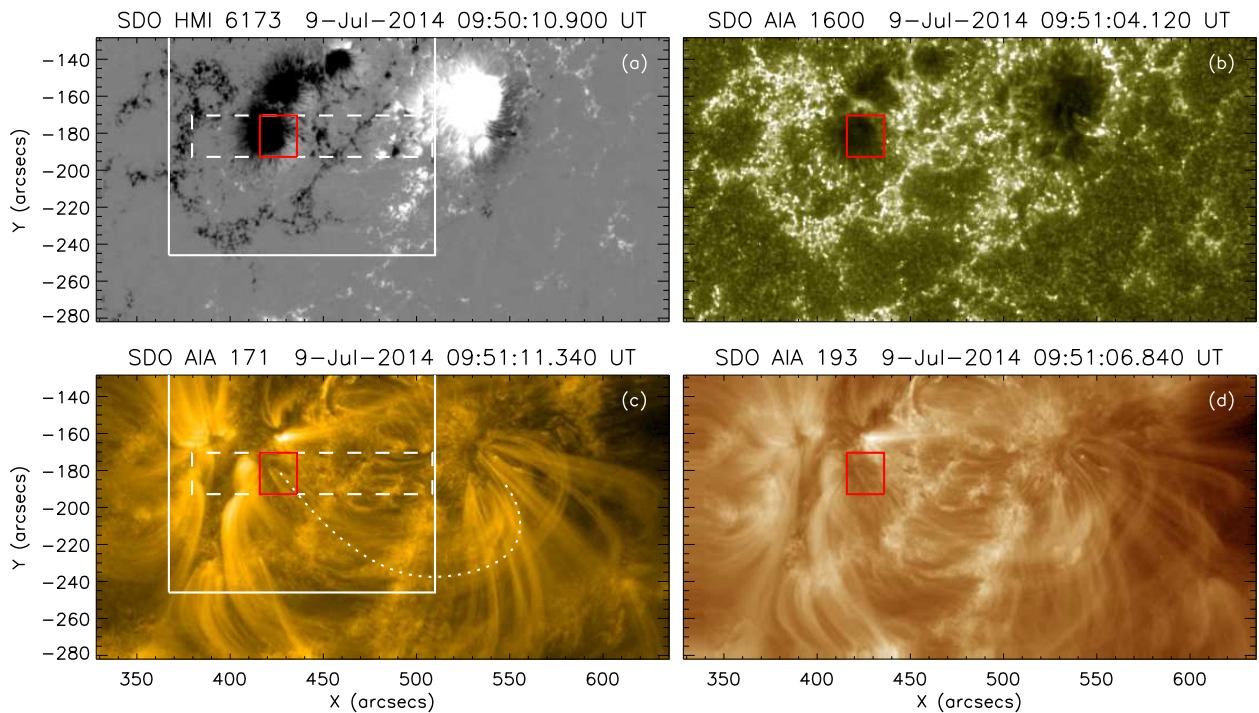}
\caption{Contextual maps of AR 12108 and the region of interest taken from the \textit{SDO}/HMI  and 
AIA observations. (a) \textit{SDO}/HMI line-of-sight magnetogram saturated at $\pm$750 G. (b) 
\textit{SDO}/AIA 1600\,\AA~channel image mainly showing the chromosphere.
Panels (c) and (d) display AIA 171\,\AA~and 193\,\AA\ maps. The intensities in panels (c) and (d) 
are in logarithmic scale. The white solid and dashed rectangles in panels (a) and (c) indicate the 
regions of IRIS slit-jaw and raster scan observations, respectively. In all the panels we identify the 
regions of interest with {isolated} umbral coronal loops in red. One such loop is traced with 
a white dotted curve. A zoom into the region indicated by the red rectangle is shown in
Fig.\,\ref{rast}. North is up.\label{context}}
\end{figure*}

There are different views based on observations whether coronal loops visible in EUV and X-ray emission connect into sunspot umbrae. For example, \citet{1975SoPh...43..327F} argued that the brightest loops in the EUV observations are associated with flux tubes leading to the umbrae of sunspots. \citet{1992ApJ...399..313S} discussed X-ray observations where they see no bright loops originating in the umbra. However, there are several studies related to bright EUV sunspot plumes and their flow channels 
\citep[e.g.,][]{1998ApJ...496L.117M,1998ApJ...502L..85B,2001SoPh..198...89B}. \citet{2008AnGeo..26.2955D} inferred that sunspot plumes are common footpoints of several active region loops. Some of the observed plumes 
probably have their roots at the centers of umbrae. Nevertheless, owing to their bright dense structure, it is often difficult to identify isolated loops in the plumes, at least close to their footpoints. In the case we report here of a rare loop being rooted in the umbra,  weak or little emission from the background umbra allows us to clearly trace the TR part of the loop to its footpoint. Furthermore, we are able to unambiguously identify these loops in the 171\,\AA\ and 193\,\AA\ channel observations obtained from the Atmospheric Imaging Assembly~\citep[AIA;][]{2012SoPh..275...17L} on board the \textit{Solar Dynamics Observatory}~\citep[\textit{SDO};][]{2012SoPh..275....3P}. These umbral loops can be completely traced to their other footpoints rooted close to the umbra of the leading opposite 
polarity sunspot.

Recently, there has been  growing interest in the detection and analysis of the evolution of small-scale 
brightenings seen in the IRIS 1400\,\AA\ slit-jaw images. \citet{2014ApJ...790L..29T} observed several 
subarcsec bright dots above sunspots and penumbra in the TR. They are mostly short-lived features (< 300 s) 
 that show broadened profiles of the Si {\sc iv} line at 1403\,\AA. Based on AIA 
observations, the authors  suggested that these dots appeared to be located at the base of loop-like 
structures. \citet{2014ApJ...789L..42K} reported bursts of supersonic downflows up to 200 km 
s$^{-1}$ and weaker upflows above a lightbridge in many spectral lines recorded by IRIS. They 
interpreted the observations as cool material falling as coronal rain along thermally unstable 
loops. They observed fast apparent downflows in the 1400\,\AA\ slit-jaw time sequence, suggesting 
that the material is indeed cooler plasma at the TR temperatures, i.e., around 10$^5$\,K. 
\citet{2015A&A...582A.116S} investigated another supersonic downflow event in a sunspot above the 
lightbridge. They identified that the downflow components are steady for almost 80 minutes and show 
up as ``satellite'' lines of Si {\sc iv} and O {\sc iv} TR lines, redshifted by $\approx$90 km 
s$^{-1}$. The authors interpreted their results as evidence of a stationary termination shock due to 
a supersonic downflow in a cool umbral loop. Such supersonic downflows in IRIS observations were 
earlier noticed by \citet{2014ApJ...786..137T}. 

In contrast to the studies of Kleint et al. (2014) and Straus et al. (2015), we investigate here the loops rooted in the dark umbra without the signatures of lightbridges. In the light of the above discussion (i.e., suppression of convective motions resulting in only a small vertical energy flux being injected into the upper atmosphere), the scenario of isolated loops in the umbra is not obvious. Furthermore, only sit-and-stare observations were used in previous studies, i.e., no detailed spatial information of \ion{Si}{iv} and \ion{O}{iv} of the structures exist. Here we  investigate maps of \ion{Si}{iv} and \ion{O}{iv} and relate them to the coronal structures seen in AIA. We study the spatial variation of the flow characteristics along the loop; in particular, we can locate and investigate the shock-front of the high-speed downflow from the upper atmosphere hitting the dense plasma at the footpoint. Having observed the footpoint and connecting loop, the results presented here will provide a better picture of the nature of flows in the TR and their possible origin.

\begin{figure*}
\centering
\includegraphics[width=17cm]{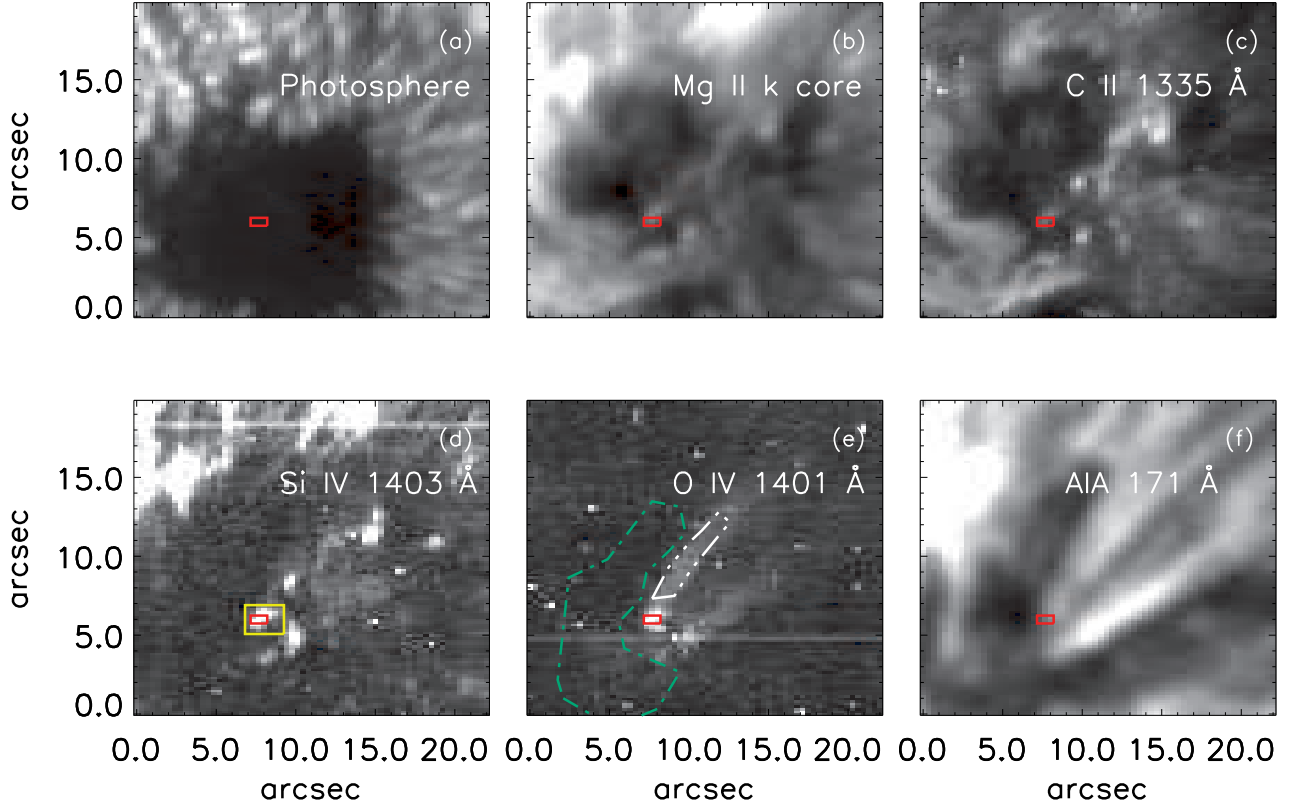}
\caption{Coronal loop rooted in sunspot umbra. The panels show the region of interest at various 
wavelengths obtained from IRIS rasters averaged over five subsequent scans (09:37 UT -- 10:05 UT).
Panel (a) shows the photosphere as seen in the far wing of \ion{Mg}{ii} near 2832.83\,\AA.
Panels (b) to (e) show the intensity in lines from the chromosphere to the upper transition region 
as recorded by IRIS and labeled with the plots. In panel (e) we mark the location of a loop 
footpoint in red, while the white region (enclosed by the dash-triple-dotted line) and green region (enclosed by the dash-dotted line) indicate the coronal loop itself and the 
background umbra, respectively. Line profiles from these regions are shown in Fig.\,\ref{avgsp}.
For comparison, the location of the footpoint is also shown in all other panels. A snapshot of the coronal 
emission as recorded by AIA in the 171\,\AA\ channel is shown in panel (f) at a time in the middle of 
the IRIS observations (09:51 UT). The field of view shown here is indicted by the red rectangle in 
Fig.\,\ref{context}c. The yellow box in panel (d) is the field of view shown in Fig.\,\ref{fits}. North is 
left (i.e., the AIA image here is rotated counter-clockwise by 90$^\circ$ with respect  to Fig.\,\ref{context}).
\label{rast}}
\end{figure*}

\section{Observations\label{obsv}}
We used calibrated level 2 data from IRIS \citep{2014SoPh..289.2733D} consisting of a time sequence of 
40 raster scans and slit-jaw images of a sunspot in the active region AR 12108 away from the 
disk center at ($423'',-182''$).%
\footnote{The data are available at http://iris.lmsal.com.}
The observations were obtained on 2014 July 9, starting at 07:21 UT. In the 
direction of the scan, each raster has 64 steps covering $22''$ with an exposure time of 4 s at 
every step and a cadence of 340 s per raster. The roll angle of IRIS slit was $90\degree$ for 
these observations. In our analysis, we mainly focus on five rasters (25--29, starting at 09:37 UT), 
which are free from contamination due to showers of charged particles on the detectors. We 
co-aligned these rasters to subpixel-level accuracy using  cross-correlations of photospheric 
scans in the wing of Mg~{\sc ii}~k. For the absolute wavelength calibration of  far-ultraviolet (FUV) spectra, we 
calculated centroids of the O~{\sc i}~line at 1355.59\,\AA. The mean of the derived line centroid was 
systematically redshifted by $\approx$ 0.056~\AA, which was subtracted from the original FUV 
wavelength array (amounting to a Doppler redshift of $\approx$ 12.3 km s$^{-1}$). In the rest of 
this paper, we follow the convention that positive velocities denote redshifts. The rest wavelengths 
of various FUV spectral lines analyzed  here are taken from~\citet{1986ApJS...61..801S}. To provide 
context to IRIS observations and to gain knowledge on the temperature diagnostics at various 
regions of interest, we have included data from six EUV channels from  AIA 
\citep{2012SoPh..275...17L}  in our analysis (94\,\AA, 131\,\AA, 171\,\AA, 193\,\AA, 
211\,\AA, and 335\,\AA). We also examined line-of-sight magnetograms from the Helioseismic and 
Magnetic Imager~\citep[HMI][]{2012SoPh..275..207S} on board \textit{SDO}.

\begin{figure*}
\centering
\includegraphics[width=17cm]{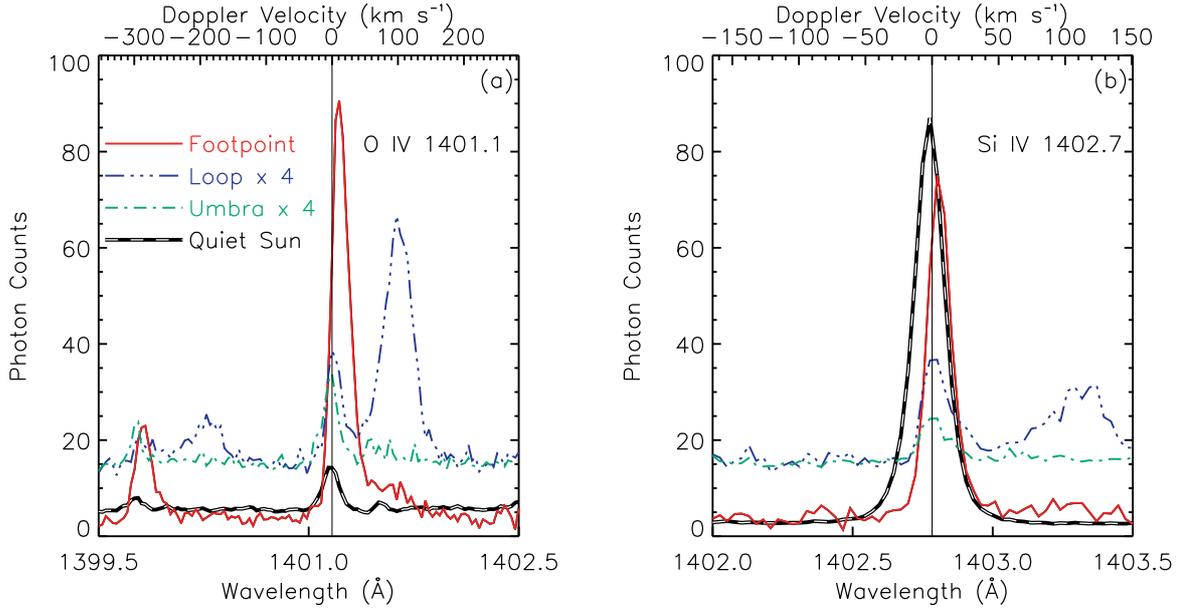}
\caption{Average line profiles of \ion{O}{iv} and \ion{Si}{iv} in selected regions.
These regions are the loop footpoint (red), the loop itself (blue) and the background umbra (green) 
as indicated in Fig.\,\ref{rast}e. The black dashed line shows an average quiet Sun spectrum 
adjacent to the sunspot (in the right panel scaled by 0.5 for better visibility). The top axes give 
the wavelength in Doppler shift units with respect to the rest wavelength of \ion{O}{iv} and 
\ion{Si}{iv}, respectively. The thin vertical line in both the panels corresponds to the 
respective zero Doppler shift. See Sect.\,\ref{S:flows}.
\label{avgsp}}
\end{figure*}

Figure~\ref{context} shows the maps of AR 12108 in various \textit{SDO} observables. The IRIS 
slit-jaw and raster scans  covered a negative polarity sunspot in that active region. In 
Fig.\,\ref{context}c AIA 171\,\AA\ loops can be seen rooted in the umbra (enclosed in the red box). 
The umbral field strength in the vicinity of these loops is in excess of 1 kG. These loops are also 
visible in AIA 193\,\AA\ (Fig.\,\ref{context}d), even though they appear less clear in this channel. 
However, they are not visible in the hotter channels of AIA (e.g., 335\,\AA). There are small, faint, bright patches at 
$(420'',-180'')$ in AIA 1600\,\AA\ channel (Fig.\,\ref{context}b), which are close to the footpoints of the umbral loops. They are most likely due to the contribution of the \ion{C}{iv} doublet at 1548\,\AA\ and 1550\,\AA\ to the 1600\,\AA\ channel because they occur at the same locations where we see bright points with IRIS in 
\ion{Si}{iv} that forms at roughly the same temperature as \ion{C}{iv} (cf.\ Fig. \ref{rast}d). By 
following the time sequence of these loops, we infer that the footpoints at the other end of 
these loops are rooted in or near the umbra of the leading sunspot with positive polarity. One such loop is 
highlighted as a white dotted curve in Fig.\,\ref{context}c.

To illustrate the overall structure of the loops and their footpoints in the umbra in Fig.\,\ref{rast} we show time-averaged  
raster maps (mean of all  five co-aligned rasters) from the  IRIS observations.  The photosphere seen at 2832.83\,\AA\ in the wing of the \ion{Mg}{ii} line showing the umbra and penumbral structures is displayed in Fig.\,\ref{rast}a. The upper atmosphere in the same field of view that was seen in the Mg {\sc ii} k core, C {\sc ii} 1335\,\AA, Si {\sc{iv}} 1403\,\AA, and O {\sc iv} 1401\,\AA\ with increasing formation temperatures of ${\log}T\,\text{[K]}=$ 4.0, 4.4, 
4.9, and 5.15 is displayed in Fig.\,\ref{rast}b--e. An AIA 171~\AA~snapshot of the same region is 
shown in Fig.\,\ref{rast}f. The cooler TR lines (C~{\sc ii} and Si~{\sc iv}) show rich structures 
in the umbra while the hotter O~{\sc iv} line shows the structures directly associated with the 
loops also seen at coronal temperatures in AIA 171\,\AA. Combined IRIS SJI at 1400\,\AA\ and AIA 
171\,\AA\ movies indicate that many of the features  are also footpoints of faint loops, but are not 
prominently seen in O~{\sc iv} line. In Fig.\,\ref{rast}e, we outline various structures (footpoint 
in red, loop in white, and background umbra in green). The footpoint and loop seen by IRIS and AIA remained spatially steady during the five raster scans, showing no lateral motions and so we estimate that the lifetime of these loops is at least 22 minutes. The position of the footpoint is overlaid on all the other panels of Fig.\,\ref{rast} to show its spatial relation with the surrounding umbra. We note that the footpoint can be seen as a small-scale bright region in the cooler lines (Fig.\,\ref{rast}b and c).

In this paper we will concentrate on one of the two loops originating from the umbra  (Fig.\,\ref{rast}e,  upper loop, enclosed by the dash-triple-dotted line in white). The other loop (lower loop), also 
clearly visible in Fig.\,\ref{rast}e shows the same properties, albeit less clearly because of a 
lower count rate and a correspondingly smaller signal-to-noise ratio in the IRIS FUV 
rasters. Although it is  fainter in the \ion{O}{iv} line that forms at 0.14 MK (Fig.\,\ref{rast}e), it appears that the lower loop is brighter than the upper loop in AIA 171\,\AA\ at 1 MK (Fig.\,\ref{rast}f). This suggests that the lower loop is  
hotter than the formation temperature of \ion{O}{iv}.


\section{Results\label{res}}
\subsection{Flows in the TR lines}\label{S:flows}
We separate the identified isolated structure into its footpoint and loop to study their individual 
properties. The spatio-temporal averaged spectra of \ion{O}{iv} and \ion{Si}{iv} in various regions  
selected in Fig.\,\ref{rast}e are plotted in Fig.\,\ref{avgsp}. 

Two prominent Doppler velocity features, which are spatially decoupled, emerge from these plots. 
 The first, the main component at the footpoint (in red), is a subsonic downflow at about 10\, km\,s$^{-1}$ to 15 
km\,s$^{-1}$ with a weak tail at higher redshifts that reaches 100\,km\,s$^{-1}$ (visible in Fig.\,\ref{avgsp}a). 
 The second, loop segment (in blue) shows a dominant supersonic downflow at 100\,km\,s$^{-1}$ with an apparent subsonic downflow component at 10\,km\,s$^{-1}$. In the \ion{O}{iv} profiles (Fig.\,\ref{avgsp}a), almost all of this slower component in the loop, at 10\,km\,s$^{-1}$, can be associated with the background emission emerging from the umbra (green). However, in the \ion{Si}{iv} loop profile (Fig.\,\ref{avgsp}b), intensity from the umbra alone cannot account for the observed slower component  because the \ion{Si}{iv} rasters of the sunspot show much richer structures in the umbra than those of \ion{O}{iv} (see Fig.\,\ref{rast}d--e). For example, the \ion{Si}{iv} bright dot near $(10'',8'')$ in Fig.\,\ref{rast}d  contributes to the slower component in the \ion{Si}{iv} loop profile. The ratio of Si~{\sc iv} 1394~\AA~and 1403~\AA~lines 
at the footpoint is found to be equal to the optically thin line ratio of 2. In both the umbra and the quiet-Sun 
spectra, there are no signs of the high-speed flows, establishing the scenario that these supersonic downflows mainly exist in the loops. 

Owing to the optically thin nature of the plasma, in the cases when the footpoint and loop appear in the line of sight, the integrated emergent profiles will show both slower and faster downflow components at the same location. In these cases, there is no clear way to decompose the source regions of different flows. In particular, the spatial location of the transition from supersonic downflows in the loops to subsonic downflows at the footpoints cannot be retrieved. This problem is even more severe with sit-and-stare observations, where there is little information along and/or across the structures. Having observed the coronal loop with an advantageous geometry, we are able to assess that the transition from supersonic to subsonic downflow happens at a projected distance of within $1''$ of the footpoint. Such a detailed observation of the flow dynamics provides crucial constraints to loop models that study stationary flows including shocks~\citep[e.g.,][]{1995A&A...300..549O} and the loop models that predict the distribution of emission along the loops, particularly in UV lines~\citep[e.g.,][]{1999PCEC...24..401O}.

One peculiarity we noticed is that 
the quiet-Sun spectra in our dataset show a net blueshift. In Fig.\,\ref{avgsp}b the quiet-Sun 
spectrum (black dashed line) is weakly blueshifted compared to the zero Doppler shift 
(thin vertical line). However, it is well known that the TR emission lines with formation 
temperature below ${\log}T\,\text{[K]}=$ 5.7 show net redshift \citep{1999ApJ...522.1148P}.

\begin{figure}
\resizebox{\hsize}{!}{\includegraphics{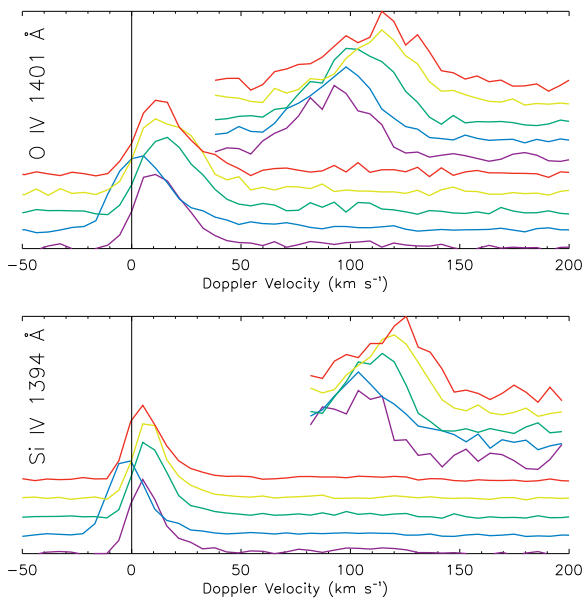}}
\caption{Stacked plot of the \ion{O}{iv} and \ion{Si}{iv} line profiles as a function of time.
The footpoint spectra cover the whole wavelength range and typically peak around  10\,km\,s\,$^{-1}$ 
redshift. The loop profiles are displayed only above 40\,km\,s\,$^{-1}$ (\ion{O}{iv}) and 
80\,km\,s\,$^{-1}$ (\ion{Si}{iv}) to avoid confusion with the footpoint spectra. They peak well 
above 100\,km\,s\,$^{-1}$.
The time the profiles are taken is color-coded (increasing from 
purple to red with an increment of 340 s and starting at 09:37 UT).
The thin vertical line indicates zero Doppler shift.
See Sect.\,\ref{S:flows}.
\label{spectim}}
\end{figure}

\begin{figure*}[!htb]
\begin{center}
\resizebox{0.33\hsize}{!}{\includegraphics{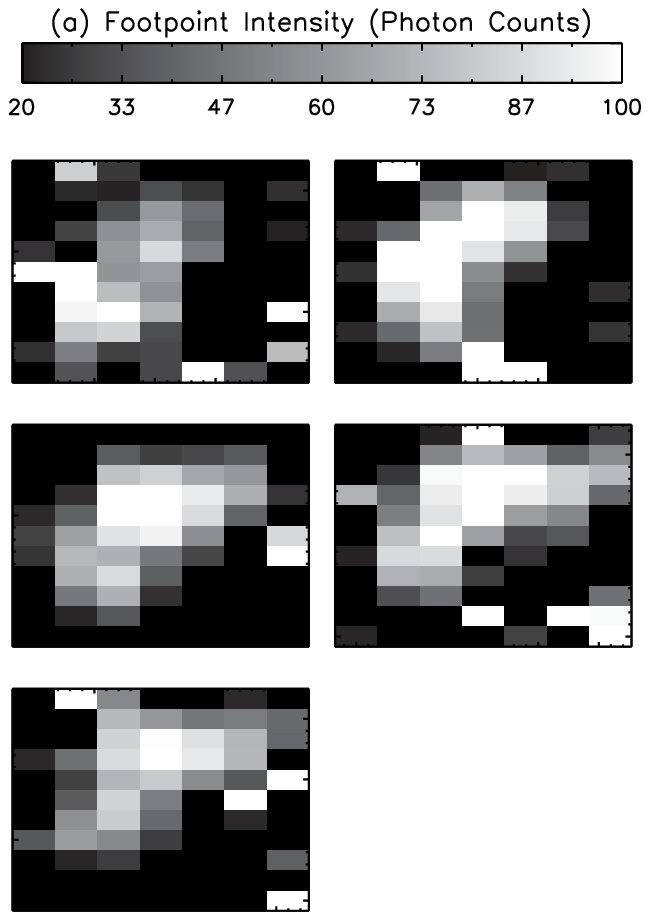}}
\resizebox{0.33\hsize}{!}{\includegraphics{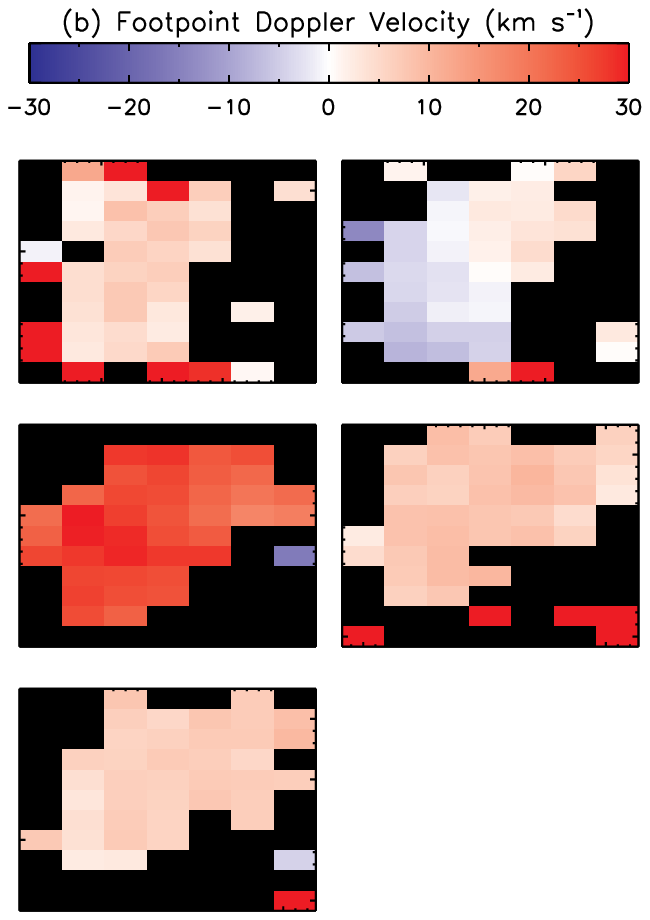}}
\resizebox{0.33\hsize}{!}{\includegraphics{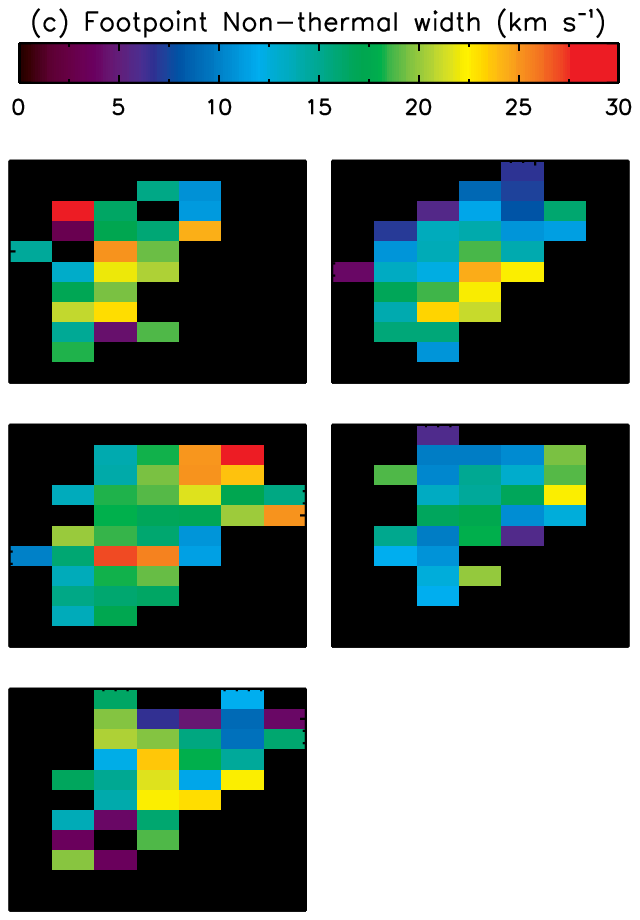}}
\caption{Maps of \ion{Si}{iv} line profile parameters.
The peak intensity (a), Doppler velocity (b), and non-thermal width (c) have been determined through 
single-Gaussian fits to the line profile of Si~{\sc iv} at each individual pixel. Each map shows a 
field of view of $\approx 2''.4 \times 1''.8$ that is indicated in Fig.\,\ref{rast}d by a yellow 
rectangle, and that fully encloses one of the footpoints of the coronal loops rooted in the umbra. 
The time the raster maps are acquired increases from from left to right and top to bottom, with a 
map cadence of about 340\,s.
See Sect.\,\ref{S:flows}.
\label{fits}} 
\end{center}
\end{figure*}

The time evolution of the \ion{O}{iv} and \ion{Si}{iv} line profiles at the footpoint and the loop 
at the TR temperatures from the five rasters are shown in Fig.\,\ref{spectim} as a stacked plot. 
The colors  indicate the time increasing from purple to red (or simply from bottom to top). The 
profiles are normalized to their respective maximum after the background subtraction. Two 
populations of spectra are plotted,   the footpoint at redshifts $\sim10\text{--}15~\text{km 
s}^{-1}$ and the loop at redshifts $\sim100~\text{km s}^{-1}$. Evidently, the footpoint shows a 
rather steady downflow compared to a gentle acceleration seen in the loop segment. This 
acceleration phase existed throughout the duration of five raster scans and thus at least 22\,minutes. The 
magnitudes of these downflows are not as large as the bursts of high Doppler shifts reported in 
\citet{2014ApJ...789L..42K}, but are comparable with the steady supersonic flows of 
\citet{2015A&A...582A.116S}. However, the latter authors find only a steady-state supersonic 
downflow, possibly associated with the lightbridge (but no loops connecting to the core of 
their umbra). Since our observations have low time resolution, we are unable to comment on any 
short-duration bursts. Whether this supersonic downflow belongs to the class of coronal rain or a
common feature in loops at the TR temperatures remains to be verified, even though an association with 
coronal rain would require a high variability, which we do not see here. In the 
coronal rain study of \citet{2014ApJ...789L..42K}, the authors found bursts of high Doppler shifts 
of several spectral lines with an average duration of about 20 s that lasted throughout the 2 hr 
 observations (cf. their Fig. 3). Observing a very steady flow at the comparably low cadence of the raster scans seems not to be consistent with the high time variability that Kleint et al. (2014) observed. If such a high variability were  present in our data we would expect a greater spatial variability along the loop in our raster maps.

Similar to \citet{2015A&A...582A.116S} we find that the supersonic downflows show a gradient in velocity between the  lines of \ion{Si}{iv} and \ion{O}{iv}. In the loop, the difference in supersonic velocities between \ion{Si}{iv} 1403\,\AA\ and \ion{O}{iv} 1401\,\AA\ lines is about $10\pm 5$ \,km\,s$^{-1}$;  the plasma downflow seen in \ion{Si}{iv} is faster. The errors quoted here are due to (1) uncertainties of $\approx1\text{--}2$\,km\,s$^{-1}$\ in measuring the respective rest wavelength of the \ion{Si}{iv} and \ion{O}{iv} lines~\citep[reference wavelength  from][cf.~\citet{2014Sci...346C.315P}]{1986ApJS...61..801S} and (2) the line-fitting uncertainties in determining the centroids of the \ion{O}{iv} and \ion{Si}{iv} line profiles. While appearing to be co-spatial in the raster maps this implies that the two lines forming at different temperatures have to originate from different source regions, so  the loop we see has to be multi-thermal. Owing to an efficient thermal conduction along the magnetic field rather than across, the coronal plasma structures are thermally insulated. This also means that the multi-thermal loop we see is probably composed of multiple strands, each at a different temperature. Explaining the different flow speeds of the two plasma components shown by \ion{Si}{iv} and \ion{O}{iv} will be a challenging task for loop modeling. At least to our knowledge no model so far predicts such a differential flow between two ions so close in temperature.

In Fig.\,\ref{fits} we show the results from single-Gaussian fits of the Si~{\sc iv} 1403\,\AA\ line at 
the footpoint. From the five rasters, we select a small region surrounding the footpoint and we 
fit a Gaussian to the line profile at each pixel. The pixel-wise footpoint Gaussian peak 
intensity maps are plotted in Fig.\,\ref{fits}a. The Doppler velocity maps (relative shifts in the peak 
centroid with respect to the rest wavelength) at the footpoint are plotted in Fig.\,\ref{fits}b. The 
footpoint non-thermal width maps (derived from Gaussian widths, using 
\verb+eis_width2velocity.pro+~routine available in \textit{solarsoft}) are shown in Fig.\,\ref{fits}c. 
Though these maps are at the limit of what IRIS can deliver, they provide some valuable 
information. The Doppler velocity distribution shows uniform subsonic downflows across the 
structure, with the  exception of an upflow patch in the second tile of Fig.\,\ref{fits}b. This upflow 
behavior  is also reflected in the footpoint spectral profiles shown in Fig.\,\ref{spectim} (second 
profile in the timeline). The origin of this upflow cannot be established with these data. The 
non-thermal widths agree closely with the previous studies 
\citep[e.g.,][]{1993SoPh..144..217D,2001A&A...374.1108P}. A statistical study of such flow 
maps at the loop footpoints will shed light on their driving mechanisms and the coronal heating process 
itself.

\subsection{Density and temperature estimates\label{dentem}}

The flow dynamics and its nature can be further scrutinized by estimating the electron 
number density ($n_\text{e}$) along the structure. To this end, we considered the O {\sc iv} 
multiplet observed by IRIS, with ratios of line pairs sensitive to the density. We 
examined two such pairs and their ratios:  the line pair 1 at 1401\,\AA\ and 1404\,\AA, and the 
line pair 2 at 1399\,\AA\ and 1404\,\AA. These lines and their ratios are in principle sensitive to 
non-Maxwellian velocity distributions~\citep{2014ApJ...780L..12D}; however, we ignore such 
conditions here as they might be less significant than the changes in density.
The \ion{O}{iv} line at 1401.16\,\AA\ has a S {\sc i} 1401.51\,\AA\ line blend, which is usually small in 
a sunspot spectrum~\citep{2002MNRAS.337..901K}. Furthermore, we assume that the S {\sc iv} 1404.77\,\AA\ 
blend at O {\sc iv} 1404.8\,\AA\ line is also very small because \cite{2001SoPh..200...91T} 
suggested that S {\sc iv} only has  a 4\% contribution to the 1404\,\AA\ feature. Noting these 
limitations, we compute $n_\text{e}$ at the footpoint ($n^{\text{fp}}_\text{e}$) and loop 
($n^{\text{loop}}_\text{e}$) from the five raster scans based on CHIANTI atomic database version 
7.1.3 \citep{1997A&AS..125..149D,2013ApJ...763...86L} for line pairs 1 and 2. The background umbra 
spectrum is subtracted from the original profiles and the ratios are obtained for the total 
intensity from the respective line profile. 

In Fig.\,\ref{den} we plot the results of derived $n^{\text{fp}}_\text{e}$ (solid) and 
$n^{\text{loop}}_\text{e}$ (dashed) from the two line pairs. 
The density $n^{\text{fp}}_\text{e}$ from pair 1 is systematically larger than that of pair 2, but 
considering the limitations of the density analysis the results from the two pairs can be considered 
 consistent (only $n^{\text{fp}}$ at 6\,min is outside the error margins). Within the error 
limits,  $n^{\text{fp}}_\text{e}$ and $n^{\text{loop}}_\text{e}$ are roughly constant in time, and 
the density at the footpoint is approximately a factor of six larger than in the loop. In general, the high-speed downflows affect the ionization-equilibrium of the plasma such that the ions are formed at temperatures different from their equilibrium temperature, which in turn affects the density estimates. However, these effects may not be very significant in our density estimates as the umbral loop is not rooted in an otherwise denser chromospheric environment (e.g., plage, network) where those effects usually become important (see Appendix~\ref{app1}).

The co-aligned EUV data from the AIA are used to construct differential emission measures (DEM) of 
the loop and footpoint with \verb+xrt_dem_iterative2.pro+ \citep{2004IAUS..223..321W,2004ASPC..325..217G}.  
The background subtracted intensity from the six AIA EUV channels (background region marked in the left panel of Fig.\,\ref{dem}), associated photon noise, and the respective filter response curves are given as input to the procedure to construct DEM (cm$^{-5}$ K$^{-1}$) at each pixel as function of temperature $T$, where $5.5\leq \text{log}_{10} ~T [\text{K}] \leq 6.5$. The results are plotted in the right panel of Fig.\,\ref{dem}. Each curve is the median of DEM solutions with $\chi^2\leq1$ obtained from a set of 300 Monte Carlo iterations. The DEMs along the loop at two locations starting from the footpoint are plotted in red and black, respectively. 

In Fig.\,\ref{dem} the DEM curve closer to the footpoint (red) shows a single peak around $\log\,T [\text{K}]{\approx} 5.7$, while the DEM curve of the loop itself (black) shows a peak at $\log\,T [\text{K}]{\approx} 6.2$. This indicates that the temperature along the loop is slowly (but not too strongly) rising from the footpoint upwards. Similar to the velocity diagnostics, the temperature estimates benefit from a clean line of sight and clear identification of the loop. In cases where the optically thin plasma at different source regions with different temperature distributions is observed in the line of sight, the two separate emission distributions seen here would merge into a single broad emission distribution with distinct peaks. Such emission measures in sunspot plumes were previously observed by e.g., \citet{1985ApJ...297..805N}. Similar distinct peaks over an underlying broad emission distribution for active region loops at the limb were observed by \citet{2008ApJ...672..674L}. Owing to the limited spatial resolution of AIA and the lack of adequate diagnostics, we cannot fully quantify how much the temperature increases along the loop.  

\subsection{Flows seen in chromospheric lines}\label{chromo}
Surrounding the loop, we also observed supersonic downflows in the Mg {\sc ii} k and h lines, which are 
cooler compared to the TR Si {\sc iv} and O {\sc iv} lines. These downflows show up as well-defined 
but broad redshifted components in the wings of the main lines at $\approx$100\,km\,s$^{-1}$. In the 
left panel of Fig.\,\ref{mgmap}, a raster map from the wing of Mg {\sc ii} k 2796.3 line averaged 
over 70\,km\,s$^{-1}$ to 120\,km\,s$^{-1}$ is shown. Bright streaks, almost co-spatial with the 
loops in Fig.\,\ref{rast}e, indicate the high-speed downflows seen in \ion{Mg}{ii}. No obvious 
footpoints in the umbra are seen in this map. These streaks are short-lived and intermittent 
structures when compared to the stable loop. They are more structured than the loops and have flows 
that go beyond the footpoint. 
In particular, the streaks seen in the \ion{Mg}{ii} map seem to have a constant cross section, as do coronal loops. Emerging directly from the umbra with a 
strongly diverging magnetic field, it is not clear at all how these constant-cross-section 
structures would form. In the right panel of Fig.\,\ref{mgmap} we plot the spectral profiles from two locations where the footpoint is observed. These locations are marked as a square (solid line) and a triangle (dashed line) in the left panel of Fig.\,\ref{mgmap}. 

By comparing the nature of two different flows, i.e., highly dynamic and structured flows in chromospheric lines and more stable downflows in the TR lines, we infer that the TR loops and the structures seen in cooler chromospheric lines have different origins and are spatially disconnected. It is puzzling that lines forming at temperatures differing by an order of magnitude still have the same supersonic redshifts. Consequently, hot and cool supersonic downflows exist, and the question of why some downflows are heated while others are not needs to be resolved. We note that \citet{2014ApJ...789L..42K} also observed enhanced red-wings of \ion{Mg}{ii} k and h lines that rapidly change their shape. At the same time, they  also noticed bursts in the TR lines. On the other hand,  
 \citet{2015A&A...582A.116S}, who observed stable downflows in the TR lines, do not detect any velocity signatures in chromospheric lines. 
 
 \section{Discussion}
 
 \subsection{Flow dynamics}
 The high-speed downflows in coronal loops can be fed by supersonic siphon flows~\citep{1980SoPh...65..251C,1995A&A...300..549O}. While other scenarios might be envisaged, we find support for the siphon flow process (see Sect.~\ref{sflow}). 
 
\begin{figure}
\resizebox{\hsize}{!}{\includegraphics{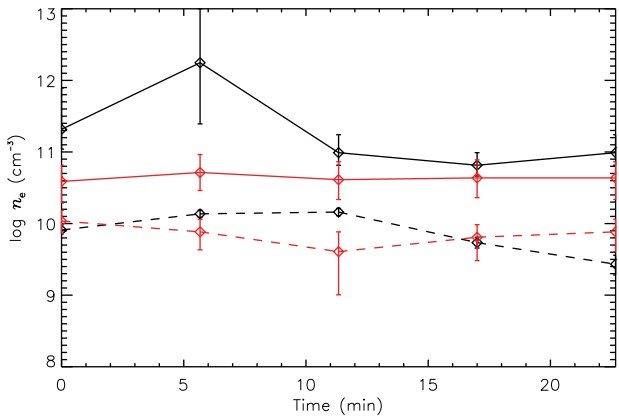}}
\caption{Density in the loop and its footpoint.
Estimated electron number densities ($n_\text{e}$) as  functions of time at the footpoint 
(solid) and loop (dashed) derived from the ratios of O {\sc iv} line pair\,1 (1401\,\AA/1404\,\AA; black) 
and line pair\,2 (1399\,\AA/1404\,\AA; red). Errors in density estimates due to standard photon noise are 
overplotted.
See Sect.\,\ref{dentem}.
\label{den}}
\end{figure}

\begin{figure*}
\sidecaption
\includegraphics[width=118mm]{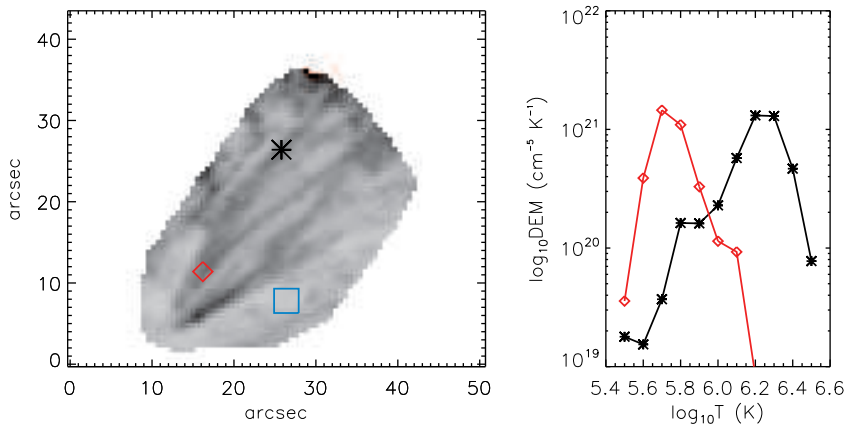}
\caption{Differential emission measure (DEM) in the loop and close to its footpoint. The left panel shows an AIA 171~\AA~snapshot (in a negative scale) with various positions marked (red: near the footpoint; black: away from the footpoint along the loop; blue: background region). In the right panel, DEMs resulting from the background subtracted data are plotted for two positions along the umbral loop. Each curve shown here is the median of DEMs with $\chi^2\leq1$ that resulted from 300 Monte Carlo iterations. See Sect.\,\ref{dentem}.
\label{dem}}
\end{figure*}

\begin{figure*}
\sidecaption
\includegraphics[width=118mm]{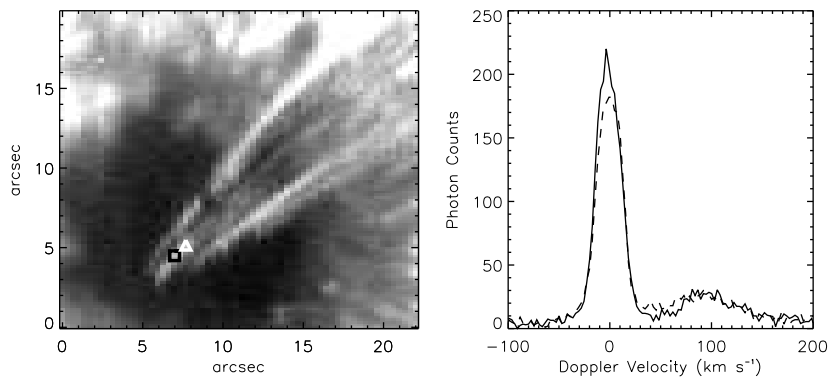}
\caption{Chromospheric emission from the loop that is rooted in the umbra.
The left panel shows a raster map (obtained from 09:37 UT to 09:43 UT) in the red wing of Mg {\sc ii} k 2796\,\AA\ line integrated from 70\,km\,s$^{-1}$ to 120\,km\,s$^{-1}$ redshift showing mainly the cool component of the downflowing plasma. The field of view is the same as in Fig.\,\ref{rast} and is indicted by the red rectangle in 
Fig.\,\ref{context}c. North is left. The right panel shows the Mg {\sc ii} k line profiles at the 
two positions marked in the left panel (the solid line corresponds to the position indicated by the 
square, the dashed line to the triangle). The wavelength is given in Doppler shift units.
See Sect.\,\ref{chromo}.
\label{mgmap}}
\end{figure*}

Supersonic downflows (including those generated due to siphon flows) have to undergo a shock transition to subsonic speeds near the footpoints. The density enhancement of six times at the footpoint (Fig.\,\ref{den}) increases the optically thin radiative losses ($\propto n_\text{e}^2$) by 36 times. It is plausible to assume that the cooling is very effective at the footpoint, such that the plasma there is isothermal with the pre-shocked plasma in the loop. In this case, we can consider that the plasma experiences an isothermal shock. The Rankine--Hugoniot condition for such isothermal shocks is given by
\begin{equation}
v_{\text{fp}}=\frac{v_\text{loop}}{M_\text{loop}^2}\label{eq1},
\end{equation}
where $v_{\text{fp}}$ ($v_{\text{loop}}$) represent the flow velocity in the footpoint (loop), and $M_\text{loop}$ is the Mach number in the loop~\citep[cf. Eq. 6 in][]{1995A&A...300..549O}. The above equation is a result of conservation of mass, momentum along with the isothermal condition in pre- and post-shocked plasma. Here we choose typical flow velocities observed in this work, i.e., $v_\text{loop}=100$\,km\,s$^{-1}$ and  $v_\text{fp}=15$\,km\,s$^{-1}$. In the loop, $M_\text{loop}=v_\text{loop}/v_s$, where $v_s$ is the sound speed in the loop. At the \ion{O}{iv} formation temperature of 0.14 MK, $v_s=43$\,km\,s$^{-1}$. The substitution of the values of $v_\text{loop}$ and $M_\text{loop}$ in Eq.\,\ref{eq1} yields $v_\text{fp}=18.5$\,km\,s$^{-1}$, which is very close to the downflows we observe in the footpoint. This suggests that the flow dynamics are consistent with mass flux conservation at a stationary isothermal shock in a constant-cross-section loop. 

In the case we presented here, the shock-front is located close to the footpoint within a projected distance of $1''$. Assessing the location of the radiative shock-front and its nature (isothermal or adiabatic) are important and challenging issues beyond the modeling of solar coronal loops. In general, radiative shocks and post-shock gas dynamics are of particular interest in the context of interstellar shocks, astrophysical jets, and their numerical models \citep[e.g.,][]{2003ApJ...591..238S,2007JCoPh.225.1427M}.

\subsection{Mass supply to sustain the downflow\label{sflow}}
Because the downflows and the related mass flux are substantial, the source of this mass feeding the high-speed downflow needs consideration. For this we examine the timescales at which plasma drains out of the loop hosting the downflow and compare it to the lifetime of the cool loop which is on the  order of at least 22 min.

The mass ($m_\text{cor}$) of plasma at coronal density $\rho_\text{cor}$ supported by a loop of volume $V$ is $m_\text{cor}=\rho_\text{cor}V$. Here $V=LA_\text{cor}$, where $L$ is the total length of the loop and $A_\text{cor}$ is its cross-sectional area in the corona. The rate of mass loss ($\dot{m}_\text{TR}$) from the loop at its base in the TR  due to a downflow with speed $v$ and density $\rho_\text{TR}$ through a surface of cross-sectional area $A_\text{TR}$ is $\dot{m}_\text{TR}=v\rho_\text{TR}A_\text{TR}$. Then the timescale $\tau$ to completely drain the plasma from the loop in the form of downflows is given by 
\begin{eqnarray}
\tau &=& m_\text{cor}/\dot{m}_\text{TR} \nonumber \\
       &=& \frac{\rho_\text{cor}}{\rho_\text{TR}}\cdot\frac{A_\text{cor}}{A_\text{TR}}\cdot\frac{L}{v}. \label{eq2}
\end{eqnarray}
The first term on the right-hand side is the density contrast between the corona and the TR. Assuming that the coronal section of the loop is at a constant pressure, this density contrast is equivalent to the temperature contrast between the 0.1 MK TR and 1 MK corona, which is $\approx0.1$. For the second term, from Fig.\,\ref{rast}e--f, it is clear that the respective widths of the loop in the TR and coronal temperatures are similar and we do not see noticeable areal expansion of the loop from its footpoint to a projected distance of at least 7 Mm in both the panels. After reaching coronal temperature, and owing to the large density scale heights in the corona, the loop  expands little, or $A_\text{cor}\approx A_\text{TR}$. As a result $\tau\approx 0.1L/v$. For a loop of 200--300 Mm long, hosting downflows of 100\,km\,s$^{-1}$, the time it takes to drain the plasma is on the order of 100 s, whereas the observed loop here exists for a much longer duration of at least 22 min.
\citet{2015A&A...582A.116S} observed steady high-speed downflows for over a duration of 80 min. Even if we make a conservative assumption that the loop expands 10 times more in the corona than in the TR~\citep[e.g., based on potential field modeling of active regions in][]{2012ApJ...746...81A}, then $\tau$ is only on the order of 1000 s.\footnote{In their study \citet{2015A&A...582A.116S} give a shorter timescale for the draining of only 10\,s. Most probably, they did not take into account that the loops can be considerably longer than the coronal pressure scale height and that the loops expand. Thus, a larger volume is available to feed the downflow, which results in the longer draining time we estimate here.}

There is not enough mass in the loop to sustain such a strong and persistent downflow. This also applies to processes such as coronal condensation~\citep{2003A&A...411..605M,2010ApJ...716..154A} because in this case the mass is ultimately   also supplied by the coronal volume in the loop (which is why the coronal condensations are highly intermittent, unlike the steady downflow that we and Straus et al. see). In this case the mass has to be supplied to the loop while it is draining on one side. If we exclude diffusion across field lines or interchange reconnection with fieldlines of  neighboring loops in the corona (and we do not see indications of these processes in the AIA data) then the only possibility is that the loop is fed from the other side through a siphon flow that is  generated by asymmetric heating.

\section{Conclusions\label{conc}}
We have presented a detailed analysis of flows, density, and temperature of an isolated coronal loop 
rooted in the center of an umbra in AR 12108. To our knowledge, this is the first clear instance
showing a comprehensive view of the isolated brightenings in the umbra, and their connection to the 
isolated coronal loops. Our results show that the footpoint and loop exhibit persistent downflows 
with dominant components at 10--15 km s$^{-1}$ and 100 km s$^{-1}$, respectively. The supersonic 
downflow component is uniform along the loop section we analyzed and has an apparent gentle 
acceleration (e.g., in \ion{O}{iv}, from 90\,km\,s$^{-1}$ to 120\,km\,s$^{-1}$ in 22 minutes). 
The density estimates at the footpoint are larger by an order of magnitude than in previous 
studies, while the temperature inferred from the DEM is consistent with earlier 
results~\citep[][]{2003A&A...406L...5D,2007PASJ...59S.727Y}. Overall, the flow dynamics suggest that 
the observed loop is in the phase of a stationary isothermal shock, probably driven by a siphon flow. 
Furthermore, we observed intermittent supersonic downflows in the red wings of cooler Mg {\sc ii} k 
and h lines at 100 km s$^{-1}$ with no identifiable roots in the umbra. These cooler downflows 
closely resemble a coronal rain scenario. We are probably looking at the elementary structures of 
sunspot plumes, which are usually very dense structures, embedded in a coronal rain environment. 

The very presence of such isolated loops in an unusual location such as the umbra, where the 
convection is believed to be suppressed, raises some interesting questions. 
Could it be that despite the strong magnetic field,  significant horizontal motions still exist in the 
umbra? In our observations we do not find evidence of umbral dots that could be associated with 
such motions, but this could  simply be due to the limited resolution and stray light. 
If the loops are not heated locally near the footpoints in the umbra, could they originate through  
asymmetric heating and the resulting siphon flows?
To understand if this is the case, a co-temporal study of the other footpoint of the loop rooted near the other sunspot would  
be needed, which is not possible in this observation because the length of the loop is much larger 
than the field of view of IRIS.
If there is no siphon flow, how can such a large mass flux be sustained for such a long period? 
Why do the supersonic downflows seem to have no projection effects and their magnitudes seem to be very 
similar irrespective of the position on the disk where they 
are observed (if previous studies are considered too)?
And finally, why is the downflow speed of about 100\,km\,s$^{-1}$ the same over a huge range of  
temperatures spanning from the chromosphere to the upper TR?
A statistical analysis of the coronal loops at the TR temperatures is imperative in order to answer these 
questions.

\begin{acknowledgements}
The authors thank the anonymous referee for the constructive comments and suggestions that helped improve the manuscript. L.P.C. acknowledges funding by the Max-Planck Princeton Center for Plasma Physics. P.R.Y. thanks the staff at the Max Planck Institute for Solar System Research for their kind hospitality. Funding for 
P.R.Y. comes from NASA grant NNX15AF48G. IRIS is a NASA small explorer mission developed and 
operated by LMSAL with mission operations executed at NASA Ames Research center and major 
contributions to downlink communications funded by ESA and the Norwegian Space Centre. \textit{SDO} 
data are the courtesy of NASA/\textit{SDO} and the AIA and HMI science teams. CHIANTI is a 
collaborative project involving George Mason University, the University of Michigan (USA), and the 
University of Cambridge (UK). This research has made use of NASA's Astrophysics Data System.
\end{acknowledgements}

\bibpunct{(}{)}{;}{a}{}{,}
\bibliographystyle{aa}

\begin{appendix}
\section{High-speed downflows and ionization\label{app1}}
The ionization and recombination timescales of several ions in the TR are much longer than the observed dynamics, in particular for \ion{Si}{iv} which is up to 100 s \citep{2006ApJ...638.1086P}, implying that the ions are formed at a range of temperatures away from their equilibrium formation temperature. In other words, the ions are said to be in a state of non-equilibrium ionization (NI). There have been several reports on the consequences of the NI effects on solar UV and EUV lines in the past~\citep[][to name a few]{1978ApJ...222..379R,1989ApJ...338.1131N,1990ApJ...355..342S}. Central to these consequences are the observed mass flows in the coronal loops, for example those generated through siphon flows due to asymmetric heating at the footpoints. \citet{1989ApJ...338.1131N} studied the effects of NI due to  such mass flows on several ionization states of carbon. They constructed coronal loop models with different loop top temperatures, densities, and flow velocities. They found that the highly ionized carbon (\ion{C}{iv} and above) significantly departs from ionization equilibrium for large flow speeds in the downflow legs of the loops (which are, however, a factor of 3--4 less than  observed here). \citet{1990ApJ...362..370S} studied the effects of NI on the solar coronal radiative losses. In particular, they found that the radiative losses are enhanced in upflowing plasma, whereas in the downflowing plasma they are suppressed. This will in turn have an impact on the coronal energy balance. \citet{1990ApJ...362..370S} also found that highly ionized oxygen (\ion{O}{v} and above) shows significant departures from ionization equilibrium.

Relevant to this study are the effects of NI on \ion{O}{iv}. Recently, \citet{2013ApJ...767...43O} conducted 3D numerical simulations of the solar atmosphere and solved the fully time-dependent rate equations for an oxygen model atom. They found that the probability density function (PDF) of \ion{O}{iv}, among other ions, is far from the expected statistical equilibrium case (at 0.14 MK). In particular, they show that the PDF is much broader and also has a significant contribution from lower temperatures  ($10^4$ K to $10^5$ K). This implies that the ions are found at lower temperatures where the densities are higher. As a result, the densities derived through line-ratios underestimate the actual densities by an order of magnitude. The earlier studies cited above also pointed to such conclusions. However, the simulations presented in \citet{2013ApJ...767...43O} are mixed polarity regions (with two dominant poles), with the chromosphere and TR being denser than a region above a sunspot. It is not clear whether their results and conclusions can be directly applied to the umbral loops reported in this work where the footpoints are rooted in an environment that is less dense than mixed polarity regions.
\end{appendix}

\end{document}